\documentstyle[psfig]{mn}

\newif\ifAMStwofonts

\ifoldfss
  \ifCUPmtlplainloaded \else
    \NewTextAlphabet{textbfit} {cmbxti10} {}
    \NewTextAlphabet{textbfss} {cmssbx10} {}
    \NewMathAlphabet{mathbfit} {cmbxti10} {} 
    \NewMathAlphabet{mathbfss} {cmssbx10} {} 
  \fi
  \ifAMStwofonts
    \ifCUPmtlplainloaded \else
      \NewSymbolFont{upmath} {eurm10}
      \NewSymbolFont{AMSa} {msam10}
      \NewMathSymbol{\upi}     {0}{upmath}{19}
      \NewMathSymbol{\umu}     {0}{upmath}{16}
      \NewMathSymbol{\upartial}{0}{upmath}{40}
      \NewMathSymbol{\leqslant}{3}{AMSa}{36}
      \NewMathSymbol{\geqslant}{3}{AMSa}{3E}

       \let\ge=\geqslant
    \fi
  \fi
\fi 

\ifnfssone
  \newmathalphabet{\mathit}
  \addtoversion{normal}{\mathit}{cmr}{m}{it}
  \addtoversion{bold}{\mathit}{cmr}{bx}{it}
  \newmathalphabet{\mathbfit} 
  \addtoversion{normal}{\mathbfit}{cmr}{bx}{it}
  \addtoversion{bold}{\mathbfit}{cmr}{bx}{it}
  \newmathalphabet{\mathbfss} 
  \addtoversion{normal}{\mathbfss}{cmss}{bx}{n}
  \addtoversion{bold}{\mathbfss}{cmss}{bx}{n}
  \ifAMStwofonts
    \ifCUPmtlplainloaded \else
      \UseAMStwoboldmath
      \makeatletter
      \new@mathgroup\upmath@group
      \define@mathgroup\mv@normal\upmath@group{eur}{m}{n}
      \define@mathgroup\mv@bold\upmath@group{eur}{b}{n}
      \edef\UPM{\hexnumber\upmath@group}
      \new@mathgroup\amsa@group
      \define@mathgroup\mv@normal\amsa@group{msa}{m}{n}
      \define@mathgroup\mv@bold\amsa@group{msa}{m}{n}
      \edef\AMSa{\hexnumber\amsa@group}
      \makeatother
      \mathchardef\upi="acc0\UPM19
      \mathchardef\umu="0\UPM16
      \mathchardef\upartial="0\UPM40
      \mathchardef\leqslant="3\AMSa36
      \mathchardef\geqslant="3\AMSa3E

       \let\ge=\geqslant
    \fi
  \fi
\fi 

\ifnfsstwo
  \DeclareMathAlphabet{\mathbfit}{OT1}{cmr}{bx}{it}
  \SetMathAlphabet\mathbfit{bold}{OT1}{cmr}{bx}{it}
  \DeclareMathAlphabet{\mathbfss}{OT1}{cmss}{bx}{n}
  \SetMathAlphabet\mathbfss{bold}{OT1}{cmss}{bx}{n}
  \ifAMStwofonts
    \ifCUPmtlplainloaded \else
      \DeclareSymbolFont{UPM}{U}{eur}{m}{n}
      \SetSymbolFont{UPM}{bold}{U}{eur}{b}{n}
      \DeclareSymbolFont{AMSa}{U}{msa}{m}{n}
      \DeclareMathSymbol{\upi}{0}{UPM}{"19}
      \DeclareMathSymbol{\umu}{0}{UPM}{"16}
      \DeclareMathSymbol{\upartial}{0}{UPM}{"40}
      \DeclareMathSymbol{\leqslant}{3}{AMSa}{"36}
      \DeclareMathSymbol{\geqslant}{3}{AMSa}{"3E}

       \let\ge=\geqslant
    \fi
  \fi
\fi 

\ifCUPmtlplainloaded \else
  \ifAMStwofonts \else 
    \def\upi{\pi}
    \def\umu{\mu}
    \def\upartial{\partial}
  \fi
\fi

\title[Power-law type spectrum of a ULX in IC 342]
{
Another Interpretation of the Power-law type Spectrum of an 
Ultraluminous Compact X-ray Source in IC 342
}
\author[A. Kubota, C. Done and K. Makishima]
       {A.~Kubota,$^{1,2}$ C.~Done$^2$ and K. Makishima$^3$ \\
	$^1$Institute of Space and Astronautical Science, 
3-1-1 Yoshinodai, Sagamihara, Kanagawa 229-8510, Japan\\
	$^2$Department of Physics, University of Durham, South Road, 
Durham, DH1 3LE\\
	$^3$Department of Physics,  University of Tokyo, 
7-3-1 Hongo, Bunkyo-ku, Tokyo 113-0033, Japan
        }
\date{Accepted 2002 * **.
      Received 2002 * **;
      in original form 2002 * **}
\pagerange{\pageref{firstpage}--\pageref{lastpage}}
\pubyear{1994}

\begin{document}

\maketitle

\label{firstpage}

\begin{abstract}

The ultraluminous compact X-ray sources (ULXs) generally show a
curving spectrum in the 0.7--10 keV {\it ASCA} bandpass, which looks like a
high temperature analogue of the disk dominated high/soft state
spectra seen in Galactic black hole binaries (BHBs) at high mass
accretion rates. Several ULXs have been seen to vary, and to make a
transition at their lowest luminosity to a spectrum which looks more like 
a power law.
These have been previously interpreted as the
analogue of the power law dominated low/hard state in Galactic BHBs.
However, the ULX 
luminosity at which the transition occurs must be at least 10--50
per cent of the Eddington limit 
assuming that their 
highest luminosity phase corresponds to the Eddington limit, 
while for the Galactic BHBs the
high/soft--low/hard transition occurs at a few per cent of the
Eddington limit. 
Here we show that the apparently power law spectrum
in a ULX in IC342 can be equally well fit 
over the {\it ASCA} bandpass
by a strongly Comptonised 
optically thick accretion disk with the maximum temperature of $\sim 1$ keV.
Recent work on the Galactic BHBs has increasingly shown that
such components are common at high mass accretion rates, and that 
this often characterises the very high (or {\it anomalous}) state.  
Thus we propose that the power law type ULX spectra are not to be
identified with the low/hard state, but rather represent the
Comptonisation dominated very high/{\it anomalous} state in the Galactic BHBs.

\end{abstract}


\section{Introduction}

Since the {\it Einstein} era, many ultraluminous compact X-ray sources
(ULXs; Makishima et al. 2000) with X-ray luminosities 
$L_{\rm X} \ge 10^{39-40}~{\rm erg~s^{-1}}$ 
have been found in spiral arms of nearby
spiral galaxies (e.g., Fabbiano 1989).  Such high values of $L_{\rm X}$
exceed the Eddington limit, $L_{\rm E}$, 
for a neutron star by several orders of
magnitude, and instead suggest massive (30--100 $M_\odot$) accreting
black holes (BHs). However, it is difficult to understand how such
massive BHs could form: single massive stars have extreme mass
loss throughout their life, and the maximum BH
mass expected is of order 10--15
$M_\odot$ (e.g. Fryer \& Kalogera 2001). Since there are no nearby
systems either in our galaxy or in M31 which could be easily studied
then the nature of the ULX remained a mystery.

{\it ASCA} (Tanaka, Inoue, \& Holt 1994) data led to a breakthrough
for these objects by providing the first moderate resolution energy
spectra.  As reported by many authors (e.g., Makishima et al. 2000,
and references therein), it is clear that the majority of the most luminous 
ULXs exhibit spectra which are
well fitted by the multicolor disk model (MCD model; Mitsuda et
al. 1984), similar to the case of Galactic/Magellanic BH binaries (BHBs) 
at high mass accretion rates, $\dot{M}$ (e.g. Makishima et al. 1986; 
the review by Tanaka \& Lewin 1995).  This, together with the 
variability of these systems 
including periodic variation (Bauer et al. 2000; Sugiho et al. 2001)
and their general association with regions of ongoing star
formation (Zezas, Georgantopoulos, \& Ward 1999; Roberts \& Warwick
2000; Fabbiano, Zezas, \& Murray 2001) has led to their
identification with BHBs. 

There are then three possibilities, first that these really are
intermediate mass BHBs, formed perhaps via mergers of massive
stars/BHs in a compact star cluster (e.g. Ebisuzaki et al.
2001). Alternatively, they could be ``normal'' mass ($\sim 10 M_\odot$) 
stellar BHs 
accreting beyond the critical accretion rate at which 
the disk luminosity $L_{\rm disk}$ 
reaches $L_{\rm E}$.
Recent work suggests that 
the Eddington limit can be violated by the disk becoming clumpy (e.g., Krolik 1998;
Gammie 1998; Begelman 2002 and references therein), and it has long been known
observationally that super critical accretion can happen (e.g. Cir X-1). 
A third alternative is that these are
``normal'' mass BHs, but that  their X-ray luminosity is
strongly anisotropic (beamed), so that the bolometric luminosity is
overestimated. This beaming is highly unlikely to be the relativistic
beaming seen in jet sources as this generally leads to a power-law (PL)
type of spectrum (e.g. blazars), very unlike the curving spectrum seen
by {\em ASCA} for the majority of the ULX.  However, strong anisotropy
of the disk flux might be produced if the disk is geometrically thick
in its inner regions.  The radiation can then be strongly collimated
by a funnel (e.g. Madau 1988), although the factor of 10--100 required
(King et al. 2001) seems extreme. Such an extreme thick disk might be
expected to form only under a super critical accretion rate. 
So this is merely an addendum to the high $\dot{M}$ scenario, 
rather than an independent alternative.

All these alternatives involve an extreme of one kind - either mass,
radiation luminosity or disk shape. One way to test these is to compare
the ULX spectra and spectral variability 
with those of Galactic BHBs. 
This has become much more feasible in recent years with the 
unprecedented volume of data from the Galactic BHBs gathered by {\it RXTE}. 
Here we use the bright Galactic BHB transient XTE~J$1550-564$ to
observationally determine the spectra and spectral variability of high
$\dot{M}$ disks, and show that the ULX are indeed compatible
with being massive (30--100 $M_\odot$) BHs 
accreting at close to the critical accretion rate. 

\section{A comparison of the ULX with BHB}
Here we specify the puzzles in understanding ULXs as 
$\sim 100~M_\odot$ BHs.  It has long been recognized that the
Galactic/Magellanic BHBs reside in either of the two distinct spectral
states, the high/soft state or the low/hard state.  In the hard
state, in which $\dot{M}$ is generally low,
the BHBs exhibit a hard PL spectrum in the 2--20 keV band. By
contrast, the soft state is generally seen when $\dot{M}$ is high, and it is
characterized by the MCD spectrum, which approximates the optically
thick standard accretion disk (Shakura \& Sunyaev 1973).  Based on
this viewpoint, the MCD-type ULXs have so far been regarded as
residing in the usual soft state, even putting aside the issue 
of how to make the required $\sim100~M_\odot$ BHs.

An increasing number of ULXs shows
spectral transitions from an MCD (curving) type of spectrum to one which is
better described by a hard PL. This has  been interpreted as
the transition to the usual hard state
(e.g. Kubota et al. 2000, hereafter Paper I; 
La Parola et al, 2001; Mizuno 2000).
However, such a straightforward analogy between ULXs and BHBs
gives the following puzzles. 
\begin{enumerate}
\item 
The ``soft state'' ULXs have inner disk temperatures $T_{\rm in}$ 
which are too high for their implied high masses.
Equivalently, when the BH mass is estimated from the Eddington argument,
$R_{\rm in}$ with reasonable correction for a boundary condition 
(Kubota et al. 1998) and spectral hardening factor 
(Shimura \& Takahara 1995)
falls much below the last stable orbit
for a non-spinning BH, $6R_{\rm g}$,
where $R_{\rm g}=GM/c^2$ is the gravitational radius.
This contrasts with soft-state BHBs, where
$R_{\rm in}$ generally  agrees with  $6R_{\rm g}$.

\item
The value of $R_{\rm in}$ is time variable,
in five bright ``soft state'' ULXs
including the two in IC 342  (Mizuno et al. 2001).
This again makes a contrast to the case of soft-state BHBs,
where $R_{\rm in}$ is approximately constant for each source.
 
\item 
Assuming that the ``soft state'' corresponds to $L_{\rm E}$,
the threshold luminosity for the spectral transition of ULXs
($\sim 0.3~L_{\rm E}$; Paper I) becomes
much higher than is seen among BHBs, typically 0.01--0.03 $L_{\rm E}$.  
Of course, the MCD-state luminosity could be much below $L_{\rm E}$,
but then the BH mass in ULXs would have to be even higher.
\end{enumerate}

Makishima et al. (2000) attribute the first puzzle to an extreme BH
rotation, because the last stable orbit is then reduced to $1.23R_{\rm g}$, 
although the other two problems remain unsolved. 
They suggested the scenario of Kerr BHs shining at $L_{\rm E}$ for ULXs. 
Following Makishima et al. (2000), Ebisawa et al. (2001) 
showed that, with standard Shakura-Sunyaev disks, 
even with Kerr BHs it is difficult to
obtain the required high temperatures without super (or near) 
critical accretion rates.
A more plausible explanation is that at high $\dot{M}$ the disk
structure changes from that of the standard disk due to the
disk being so optically thick that radial advection of the radiation
becomes important.  These disks, called ``slim disk'' 
(Abramowicz et al. 1988),
are different in structure from the
standard disks. Pressure support becomes important, so the material
is not in Keplarian rotation. 
The inner edge of the optically thick accretion disk 
is then not necessarily at the last stable orbit,
but can be closer to the BH
(e.g. Abramowicz et al. 1988; Watarai et al. 2000).
The small and 
{\em changing} inner disk radius (puzzles i and ii) could be explained in a
Schwarzschild metric if the disk penetrates increasingly further into
the plunging region at high $\dot{M}$ rather than abruptly
truncating at the last stable orbit, so
Mizuno et al. (2001) and Watarai, Mizuno \& Mineshige (2001) 
proposed that MCD-type
ULXs may have {\it slim accretion disks} rather than standard ones.

The inner disk radius, however,  
is very dependent on the viscosity
prescription. For small viscosity it can decrease to $\sim 4 R_{\rm g}$ for
super Eddington luminosities, whereas for large viscosity it can be
slightly larger than $6 R_{\rm g}$ (e.g. Abramowicz et al. 1988; Artemova et
al., 2001).  Also, this mechanism for producing a decreasing radii can
only be used for Schwarzschild BHs as in extreme Kerr BHs there
is so little space between the last stable orbit and the horizon.
The spectra of these slim disks have been calculated by Watarai \&
Mineshige (2001) (Schwarzschild, low viscosity) and Beloborodov (1998)
(Schwarzschild and Kerr, high viscosity). Since some of the emission
from the smallest radii is advected rather than radiated, the spectra
have {\sl less} high temperature emission than a standard disk, but
this can be somewhat compensated by the decrease in inner radius for
the Schwarzschild, low viscosity disks.  
Since the standard disk
structure calculations have difficultly in producing the high
temperatures observed from the ULX even with Kerr BHs for sub-critical 
rates (Ebisawa et al. 2001), 
then it seems unlikely that 
the slim disks will substantially help problems (i) and (ii) unless
the accretion rate is super-critical. 

Therefore it seems likely that the standard curving MCD-type ULX spectra are from 
super critical accretion. In which case the third puzzle becomes
acute. Recently, Kubota et al. (2001a,b) has observationally suggested a
novel understanding of high-luminosity accretion disks, 
using {\it RXTE} 
data on some Galactic BHBs. 
They have identified three distinct spectral 
{\it regimes} of the optically-thick accretion disk in XTE J$1550-564$ and 
GRO J$1655-40$.
Figure 1 shows their result on XTE J$1550-564$ (the distance and 
inclination angle are assumed to be 5~kpc and $60\degr$, respectively).

\begin{enumerate}
\renewcommand{\theenumi}{(\arabic{enumi})}
\item The {\it standard regime} where $R_{\rm in}$ remains constant when fit with MCD models, i.e. $L_{\rm disk}\propto T_{\rm in}^4$.
\item The luminous {\it apparently standard regime} 
where the disk luminosity rises slightly
less quickly with temperature, 
$L_{\rm disk}\propto T_{\rm in}^{\sim 2}$, 
and the spectral shape is slightly distorted from the standard-disk
one. In the literature, 
both this and (1) are identified as the soft state, with disk
dominated spectra.
\item The intermediate {\it anomalous regime}, 
where the disk Comptonisation suddenly increases and
the spectrum becomes much harder. This is often termed the very high
state in the literature 
(Miyamoto et al. 1991). 
\end{enumerate}
Kubota et al. (2001a,b) suggested a picture for this, in which the
{\it standard regime} is where the accretion flow is described by the
Shakura--Sunyaev geometrically thin disk approximation. As the disk
luminosity increases, this approximation breaks down, and the disk
becomes slim.  This theoretical picture explains both the change in
spectral shape (as the hotter, inner disk emission is preferentially
advected rather than radiated) and the different $T_{\rm in}$--$L_{\rm disk}$ 
relationship (Watarai et al. 2000) 
seen in the {\it apparently standard regime}, 
although for the likely mass/distance of XTE J$1550-564$
then the transition threshold
appears at $0.1 L_{\rm E}$, rather than at the predicted $L_{\rm E}$.
It should be noted that, though Comptonisation in the disk 
can shift the apparent disk temperature/radius as
measured by MCD models,
the effect of this is {\sl less} important at
high $\dot{M}$ than at low $\dot{M}$ (Merloni, Fabian \& Ross 2000), 
opposite to the observed behaviour on the $T_{\rm in}$-$L_{\rm disk}$ plane.

\begin{figure}
\centerline{\psfig{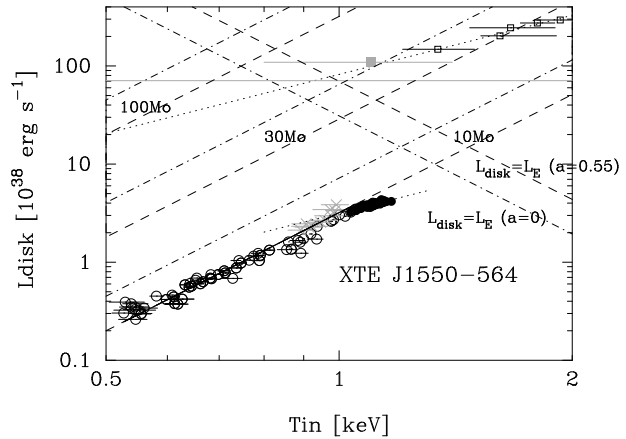}}
\caption{The calculated $L_{\rm disk}$ plotted against the observed
$T_{\rm in}$. As for XTE~J$1550-564$ (taken from Kubota 2001a), 
three kinds of symbols specify the data
obtained in the {\it standard regime} (open circles), 
the {\it apparently standard regime} (filled circles), and
the {\it anomalous regime} (gray crosses). For the {\it anomalous regime}
data, $L_{\rm disk}+L_{\rm thc}$ are plotted instead of $L_{\rm disk}$. 
The results of IC 342 source 1 in 1993, the MCD-type state, 
are also plotted with open squares.
For comparison, the 0.7--10 keV absorption corrected luminosity  
of the same source
in the PL-type state is presented as a horizontal solid line. 
The bolometric luminosity calculated under the 
{\it thcomp} model is also plotted against $T_{\rm in}$ 
with a filled gray square.
The dash-dotted lines represent the
standard $L_{\rm disk}\propto T_{\rm in}^4$ relation for 
Schwarzshild BHs of 10, 30, and 100 $M_\odot$, while 
the dashed lines represent the same relation for 
mildly rotating BHs ($a=0.55$).
Relativistic effects are not taken into account to calculate these lines. 
The dotted lines represent 
$L_{\rm disk}\propto T_{\rm in}^2$ relation.
The Eddington limits are also shown for Shwarzshild BHs and 
mild Kerr BHs with a dash-dotted line and a dashed line, respectively.
}
\label{fig:t-l}
\end{figure}

The clear observational result is that the high $\dot{M}$ 
Galactic BHBs show a variety of spectral
shapes, which correlate fairly well with $L_{\rm disk}/L_{\rm E}$. 
Overlaid on the BHB data in Fig.~\ref{fig:t-l} 
are the results from fitting several
spectra from the ULX IC~342 source 1 with the MCD model (as in
Mizuno et al. 2001, except that an inclination angle of
$60\degr$ is assumed). 
We thus notice a clear similarity between the 
ULX behaviour and that of XTE J$1550-564$ in its 
{\it apparently standard regime}. This similarity argues for
a slim disk interpretation of ULX 
spectra, as first suggested by Mizuno et al. (2001).
As to the MCD-type spectrum of ULXs, we here simply point out 
its similarity to the 
{\it apparently standard regime} of BHBs,
although to get the same temperature from a much more
massive BH requires that the disk should be accreting at a
correspondingly larger fraction of the critical rate.
Overlaid on Fig.1 is the standard
disk $T_{\rm in}$-$L_{\rm disk}$ relation for 10, 30 and 100 solar mass 
Schwarzchild and Kerr (a=0.55) black holes.

In Fig.1, the horizontal 
solid line indicates the 0.7--10 keV luminosity of IC 342
source 1 in 2000, 
when it made a spectral transition into the 
PL-type spectrum. The 
extrapolated 
1--30 keV luminosity in the PL-type spectrum is close to $L_{\rm disk}$ in the 
dimmest MCD spectral state (the 1993 data). 
This illustrates the problem (iii) above; if we require that the ULX is
shining at close to $L_{\rm E}$, then this PL-type state also has a
luminosity close to $L_{\rm E}$, so it is 
{\em unlike} a usual low/hard state for the stellar BHBs, even though
it has a characteristic PL spectral shape of the low/hard state. 

Here we propose instead that the PL-type spectrum seen in the ULX
marks a transition to the Comptonised {\it anomalous regime} rather than the 
usual low/hard state. 

\section{Reanalysis of PL-type spectrum}
In these few years,
our understanding of both ULXs and stellar BHBs 
has thus made rapid progress observationally and theoretically.
This condition makes us possible to re-consider
the PL-type ULXs and solve the third puzzle presented in \S 2.
Accordingly, we re-examine the GIS data of IC 342 source 1 in 2000, 
the longest observation among the PL-type ULXs.
The observational details are given in Paper I and Sugiho et al. (2001).

\subsection{Characteristics of the spectrum}
\begin{figure}
\centerline{\psfig{file=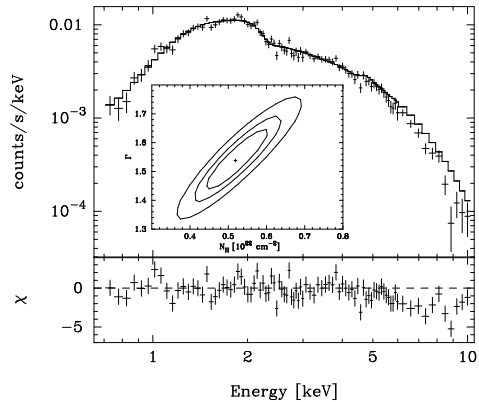,height=150pt}}
\caption{The GIS spectrum of IC 342 source 1 in 2000, the PL-type
state, with a prediction of the best-fit model obtained 
by 0.7--4 keV spectral fitting.
A confidence contour of $\Gamma$ against $N_{\rm H}$ is also shown.
The bottom panel indicate the fit residuals.
}
\label{fig:spec07-4}
\end{figure}

The overall GIS spectrum is well described by an absorbed PL
of $\Gamma=1.73\pm0.06$ (Paper I).
In order to investigate whether this is truly a
simple PL shape, we fit the low and high energy band 
separately. The 0.7--4 keV GIS data give 
a flatter power law index, $\Gamma=1.54\pm0.12$, 
modified by a low energy absorption of 
$N_{\rm H}=5.2\pm0.3\times 10^{21}~{\rm cm^{-1}}$,
than that obtained
from fitting the 5-10 keV data with a single PL
where $\Gamma=2.4\pm0.3$. 
Figure~\ref{fig:spec07-4} shows the extrapolation of the 0.7--4 keV
spectral fit to the full GIS energy band, clearly showing that the 
spectrum is slightly curved. 
This could be due to spectral complexity around the iron line and edge
energies as suggested in Paper I. However, 
the low/hard state PL-spectra of galactic BHB usually show 
an opposite curvature, with the 5-10 keV spectrum going {\it harder}
than the 0.7--4 keV spectrum due to 
frequently observed soft excess 
softening the low
energy spectrum, and some reflection hardening the higher energy
bandpass. 

\begin{figure}
\centerline{\psfig{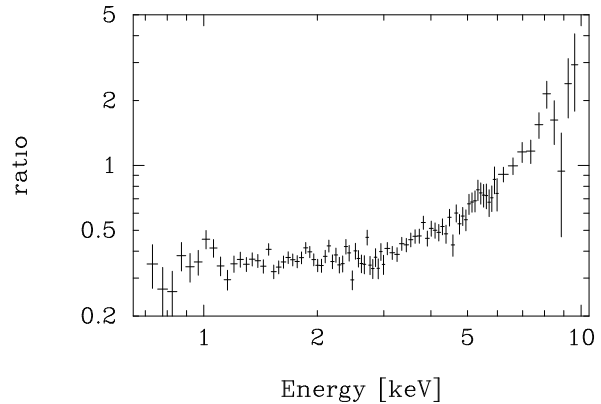}}
\caption{The PL-type spectrum of IC 342 source 1 (in 2000) 
normalized to the best-fit spectrum of the same source 
in the MCD-type state (in 1993).
The latter is convolved with the instrumental response 
calculated for the position of the source in 2000.}
\label{fig:spec_ra}
\end{figure}

Thus the subtle downwards curvature of the spectrum makes it look 
different from the BHB low/hard state.  In order to characterize the spectral
shape from another viewpoint, we compare the PL-type spectrum in 2000
with the faintest MCD-type spectra seen from the same
source in 1993 (using the same data selection as Mizuno et al. 2001).
As shown in table~1, the faintest 1993 GIS data is well fit 
with the MCD model of $T_{\rm in}=1.3$ keV modified by low energy 
absorption column of $N_{\rm H}=3.4\times 10^{21}~{\rm cm^{-2}}$.
These parameters are consistent with those in
Mizuno et al. (2001). Figure~\ref{fig:spec_ra} shows the ratio of the PL 
data in 2000 with the best fit MCD model to the 1993 faintest data. This is
remarkably flat below 3 keV, showing that the shape of the low energy
spectrum does not change. A better explanation of the PL-type spectrum could 
be that it is {\em not} a simple PL but 
has a soft component which is a fainter version of that in the faintest MCD
state, together with a Comptonised tail, similar to the {\em anomalous} 
type (or very high state) 
spectra seen in XTE J$1550-564$ and GRO J$1655-40$.

\subsection{Fit with the Comptonised model}

We re-fit the same PL-type GIS spectrum incorporating
Comptonisation. We use the same model as for XTE J$1550-564$ in the
{\it anomalous regime} (Kubota 2001a), i.e. 
an approximate model for thermal Comptonisation based on the 
Kompaneets equation 
({\it thcomp}: Zdziarski Johnson \& Magdziarz 1996, Zycki, Done, \& Smith 1999).
We choose a disk blackbody as the seed photon distribution, and
include a separate uncomptonised 
MCD whose temperature is tied to that of the seed
photons, as the {\it thcomp} model only calculates Comptonised component.
We do not include reflection as the statistics are not sufficient to
constrain the fit. 
The fit parameters are maximum temperature of the disk blackbody seed
photon spectrum, $T_{\rm in}$, the electron temperature $T_{\rm e}$,
photon index $\Gamma ^{\rm th}$ which expresses spectral shape of the
{\it thcomp} below $T_{\rm e}$, and the normalizations of the
Comptonised and uncomptonised disk photons.  We could not constrain
the values of $T_{\rm e}$ with the 0.7--10 keV GIS data.  Therefore,
we fix $T_{\rm e}$ at 20 keV as seen in XTE J$1550-564$ 
(Wilson \& Done 2001).

We show the best fit {\it thcomp} parameters in table 1. 
The {\it thcomp} model with a low energy absorption can successfully
reproduce the observed spectrum, with $\chi^2/{\rm dof}=102/86$.
The obtained value of $T_{\rm in}=1.1$ keV is slightly below the
temperature of 1.3 keV seen in the dimmest MCD-type spectrum in 1993.
Although the direct MCD component was added to the spectral model,
the data do not require it with a
90 \% upper limit to the direct disk emission being 
less than 25 \% of the total 
0.7--10 keV energy flux. The lack of a direct MCD component is consistent with 
the optical depth of $\tau\sim 3$ inferred from the 
$\Gamma^{\rm th}$ and (fixed) $T_{\rm e}$.

\begin{table*}
 \centering
 \begin{minipage}{140mm}
  \caption{Spectral parameters of IC 342 source 1
with 90\% confidence limits}
\begin{tabular}{@{}lcccccrr@{}}
\hline\hline
epoch&model&range(keV)  &$N_{\rm H}$\footnote{Column density for
absorption assuming solar abundances, in units of $10^{21}~{\rm cm^{-2}}$} &$\Gamma$ or $\Gamma^{\rm
th}$&$T_{\rm in}$ (keV)&flux\footnote{The 0.7--10 keV source flux at the top
of the atmosphere, in units of $10^{-12}~{\rm erg~s^{-1}~cm^{-2}}$}     &$\chi ^2/$dof  \\
\hline
2000&PL &0.7--4 & $5.2\pm0.3$ &$1.54\pm0.12$&---&4.0&71.7/57\\
        &PL     &5--10  &---    &$2.4 \pm0.3$   &---&6.1&14.4/19\\
\cline{2-8}
        &{\it thcomp}&0.7--10 &$3.2\pm0.4$&$2.2\pm0.4$&$1.1\pm0.3$    &       3.3&102.2/86\\
        &PL\footnote{Taken from Paper I. An ionized Fe-K edge at $8.4\pm0.3$ keV is applied, with an optical depth of $0.9\pm0.5$.}&0.7--10 &$6.4\pm0.7$&$1.73\pm0.06$       &---    &       3.7&101.1/86\\
        &PL\footnote{Fit result by a single PL with absorption column.} 
&0.7--10 &$6.8\pm0.7$&$1.78\pm0.06$       &---    &       3.9&119.4/88\\
        &MCD\footnote{Taken from Paper I}&0.7--10 &$1.9\pm0.4$&---&$2.06\pm0.08$  &2.4&120.5/88\\
\hline
1993 (faintest phase) &MCD        &0.7--10 & $3.4^{+2.1}_{-1.8}$ &---&$1.30^{+0.19}_{-0.16}$&6.4 &23.1/33\\
\hline
\end{tabular}
\end{minipage}
\end{table*}

\section{Discussion}

In \S3, we have shown that 
the apparently PL-type spectra seen from IC342 in
2000 is most probably {\em not} related to the low/hard spectra in
Galactic BHBs. It shows significant deviations from a PL shape,
in the sense that the spectrum softens at higher energies. 
This is opposite to the behaviour of the Galactic BHBs in the low/hard
state. 
We propose that the PL-type spectra are instead the analogue of the 
{\it anomalous regime} (also termed the very high state) spectra seen in
the Galactic BHs at high luminosities. 
We demonstrate this by fitting Comptonisation models, {\it thcomp}, 
to the PL-type
data, and show that they can indeed give as good a fit to the data as
a PL continuum plus ionized edge (Paper I). 
If the ionized edge is not incorporated, the PL fit is significantly 
inferior to the {\it thcomp} fit (table 1), because of the 
intrinsic concaveness of the spectral shape.

Additionally, validity of this interpretation can be tested 
by investigating the location of the obtained result on 
the $L_{\rm disk}$-$T_{\rm in}$ plane (Fig.1). 
There, we also show the 0.1--100 keV {\it thcomp} luminosity, 
$L_{\rm thc}=1.1\times10^{40}~{\rm erg~s^{-1}~cm^{-2}}$, 
extrapolated from the best fit model 
assuming isotropic emission. If the Comptonisation is from overheating
of the disk (Beloborodov 1998) then this represents the disk
luminosity. If instead it is from a separate corona, then the disk
luminosity is amplified by Comptonisation. However, at this
low compton $y$ parameter the amplification is rather low (less than a
factor of 2), so we use the {\it thcomp} luminosity
as our estimate for the
disk luminosity and plot this with the seed photon 
temperature on Fig~\ref{fig:t-l}. The PL-type spectrum then lies nicely on 
the same luminosity-temperature relation as defined by the MCD-type spectra. 
This predicts a ULX mass between $30~M_\odot$ ($a=0$) and 
$150~M_\odot$ ($a=0.998$) 
if the PL-type spectrum 
marks the break between the {\it standard} and 
{\it apparently-standard} (slim disk) regimes.
This break then appears at $\sim 1$--$2 L_{\rm E}$. If instead 
the break is at $\sim 0.1$--$0.2L_{\rm E}$ as in J1550 
then this implies a mass of 50--250 $M_\odot$.

Based on these observational results, including spectral softening 
at higher energies, X-ray luminosity, and disk temperature within the 
framework of the disk Comptonisation,  
we conclude that 
the PL-type spectra of IC 342 source 1 in 2000
are related to the standard MCD-type 
disk with strong disk Comptonisation state rather than to the low/hard state, 
and that the transition between 1993 and 2000 is likely an 
{\it apparently standard} to {\it anomalous} transition, 
rather than the canonical high/soft to low/hard transition.

\subsection*{Acknowledgements}

A.K. is supported by Japan Society for the promotion of Science
Postdoctoral Fellowship for Young Scientists.
The present work is supported in part by 
Sydney Holgate fellowship in Grey College, University of Durham.



\end{document}